\documentclass[preprint,aps,floatfix,showpacs,a4paper]{revtex4-1}

\usepackage{amsmath,amssymb,amsfonts,dcolumn,eulervm,color,graphicx,graphics,latexsym,placeins,epsfig,subfigure,hyperref}
\usepackage{amsmath}

\newcommand{\be}{\begin{equation}}
\newcommand{\ee}{\end{equation}}

\newcommand{\lb}{\left}
\newcommand{\rb}{\right}

\newcommand{\arxiv}[1]{\href{http://arxiv.org/abs/#1}{arXiv:#1}}    

\definecolor{darkred}{rgb}{.8,0,0}

\definecolor{darkblu}{rgb}{0,0,.8}

\definecolor{darkgreen}{rgb}{0,.8,0}

\begin{document}

\title{Thermodynamical universality of the Lovelock black holes}

\author{Naresh Dadhich}\affiliation{ Inter-University Centre for Astronomy \& Astrophysics,\\ Post Bag 4, Pune 411 007, India}\email{nkd@iucaa.ernet.in}

\author{Josep M. Pons}\affiliation{ DECM and ICC, Facultat de F\'{\i}sica, Universitat de Barcelona,\\ Diagonal 647, 08028 Barcelona, Catalonia, Spain.}\email{pons@ecm.ub.es}

\author{Kartik Prabhu}\affiliation{Department of Physics, University of Chicago,\\5640 S. Ellis Avenue, Chicago, IL 60637, USA}\email{kartikp@uchicago.edu}

\begin{abstract}
The necessary and sufficient condition for the thermodynamical
universality of the static spherically symmetric Lovelock  black
hole is that it is the pure Lovelock $\Lambda$-vacuum  solution.
By universality we mean the thermodynamical parameters:
temperature and entropy always bear the same relationship to the
horizon radius irrespective of the Lovelock order and
the spacetime dimension. For instance, the entropy always goes in terms of the
horizon radius as $r_h$ and $r_h^2$ respectively for odd and even
dimensions. This universality uniquely identifies the pure Lovelock black
hole with $\Lambda$.
\end{abstract}


\maketitle

Black hole and big-bang singularity are the two profound defining
predictions of Einstein's gravity and they are therefore also the
test markers for its generalizations and extensions. One of the
obvious questions that arises is what happens if we go to higher
dimensions, is it the same Einstein-Hilbert Lagrangian or does it
need to be generalized? What should be required of the generalized
Lagrangian? It is natural to ask for (a) general covariance - a
 scalar density constructed from the Riemann curvature
which yields a non-trivial equation in a given dimension, (b) the
equivalence principle, and (c) the equation of motion to be second
order quasi-linear. This uniquely identifies the Lanczos-Lovelock
Lagrangian (LL-gravity) which is a homogeneous polynomial in the
Riemann curvature with specific coefficients where zeroth, linear
and quadratic orders respectively correspond to $\Lambda$,
Einstein-Hilbert and Gauss-Bonnet \cite{lovelock}. It turns out
that such higher order terms occur in the low-energy effective
action in string theory \cite{string}. It has also been argued
that the inclusion of such higher order terms in the gravitational
action is indeed motivated purely on classical considerations for
incorporation of high energy effects \cite{dad}. It is clear that
it is the requirement (c) that ensures the unique physical
evolution for a given initial value problem and it is also
tantamount to the characterization of the Lovelock action by the
Bianchi derivative \cite{Dadhich:2008df} and by the Levi-Civita consistent truncation \cite{Dadhich:2010}. The LL-gravity is
therefore the most natural generalization of the  Einstein gravity
in the strong gravitational regime where higher order curvature
terms may become important and represent high energy corrections. \\

For probing gravitational dynamics in higher dimensions, we shall
employ the study of the Lovelock black holes in higher dimensions.
There is a very extensive body of work on this topic beginning with
the two classic papers \cite{wheeler,whitt}. For static spherically
symmetric vacuum solutions, the equation ultimately reduces to a
first order differential equation involving an $N$-th order
algebraic polynomial. The critical issue is therefore to crack this
polynomial. As is well known there exists no standard method to
solve it for the order $N>4$. Besides there is a problem of
extracting meaningful physical information from black hole in terms
of its thermodynamical parameters like temperature and entropy which
become totally unmanageable if there are too many coupling
parameters involved. It is therefore pertinent to restrict the
number of couplings and the natural choice is one the zeroth order
Lovelock, $\Lambda$ which is a constant of spacetime structure like
$c$ \cite{dad-const} and the other is the gravitational constant
(Planck mass) for any Lovelock order - the usual $G$ for the first
order Lovelock, Einstein-Hilbert. But many other simplifying choices
are possible for circumventing this situation. For instance one
could take the polynomial to be completely degenerate which is then
trivially solvable \cite{kpd}. This would mean that all Lovelock
coefficients are not independent but are given in terms of the
single one, $\Lambda$, and which would also
define a unique $\Lambda$-vacuum. \\

Interestingly this is exactly the ansatz arrived at for the
dimensionally continued black holes \cite{BTZ,CTZ} from entirely
different considerations of the continued extension of the Euler
density to next dimension, the embedding of Lorentz group
$SO(d-1,1)$ into the larger AdS group $SO(d-1,2)$ and also to have
the unique value of $\Lambda$. We would like to motivate our ansatz
by asking the universal thermodynamical behavior for black holes.
That the thermodynamical parameters always bear the same functional
form with the horizon radius for odd $d=2N+1$ and even $d=2(N+1)$
dimensions irrespective of the Lovelock order $N$. For instance the
entropy always goes as $r_h$ and $r_h^2$ respectively for odd and
even dimensions. Thus black hole thermodynamics is not at all
sensitive to the Lovelock order. This physical requirement uniquely
identifies the pure Lovelock black hole with $\Lambda$; i.e. it is
the solution of the $N$th order pure Lovelock $\Lambda$-vacuum
equation \cite {LL} (henceforth the pure Lovelock with $\Lambda$
would simply be called the pure Lovelock black hole). It is both a
necessary and sufficient condition; i.e. the pure Lovelock black
holes have universal thermodynamics and the universality of
thermodynamics uniquely characterizes the pure Lovelock black holes.
Like the dimensional continuity is identified with the degeneracy of
the algebraic polynomial, it is the derivative
degeneracy that includes the pure Lovelock black holes. \\


Thus the main motivation of this letter is first to expose the
thermodynamical universality of the pure Lovelock black hole and
then show that it is its unique characterization. This is a very
remarkable general property of the higher dimensional Lovelock
gravity. The first time such a universality was discovered for
gravity in higher dimensions was for the case of uniform density
fluid sphere which was shown to be always given by the
Schwarzschild interior solution irrespective of whether it is
Einstein or Einstein-Lovelock gravity of any order in higher
dimensions \cite{dmk}. That is gravity inside a uniform density
sphere has universal character, it doesn't matter whether it is
described by the Newtonian, Einsteinian or in general Lovelock
gravity. There also exists universality of the large $r$ behavior
of the pure Lovelock as well as Einstein-Lovelock black holes.
That is all solutions asymptotically tend to the corresponding
Einstein solution \cite{LL}. This is what it should be because
higher order curvature contributions through the Lovelock action
should contribute non-trivially only at the high energy end for
$r\to r_h$ while they should all wean out at the low energy
$r\to\infty$ limit. \\

We shall prove in the following that the thermodynamical universality
is the necessary and sufficient condition for the pure Lovelock black hole.\\

Let us begin with the static spherically symmetric metric and seek
solution of the vacuum equation which would in general be given by
$\sum\alpha_i\mathcal G_{ab}^{(i)} = 0$ where $\alpha_0=\Lambda,
\mathcal G_{ab}^{(0)}=g_{ab}$, $\mathcal G_{ab}^{(1)}=G_{ab}$ is
the Einstein tensor, $\mathcal G_{ab}^{(2)}$ is the quadratic
Gauss-Bonnet analogue, and so on. For the pure Lovelock solution,
it is $\Lambda g_{ab} + \mathcal G_{ab}^{(N)} =0$ and the solution
is then given by \cite {LL},
\be
V(r) = -g_{tt} =g_{rr}^{-1} = 1-r^2\lb(\alpha_0 + \frac{\mu}{r^{d-1}}\rb)^{1/N}
\label{thepot}
\ee
where $\mu$ isthe black hole mass parameter. The black hole temperature and
entropy are readily computed by evaluating the expressions $T  =
\frac{1}{4\pi} V'(r_h ), ~\, S  = \int T^{-1}d\mu = \int_0^{r_h}
T^{-1}\frac{\partial \mu}{\partial r_h} dr_h$ and so we obtain
\begin{subequations}
    \begin{align}\label{eq:LL T}
        T   = \begin{cases}
               \frac{\mu -1}{2\pi r_h} & d = odd    \\[10pt]
              \frac{1}{2\pi}\lb[- \frac{1}{r_h} + \frac{d-1}{d-2}~\frac{\mu}{r_h^2}\rb] & d = even  \\
          \end{cases}\\[15pt]
            \label{eq:LL S}
        S  = \begin{cases}
                    2\pi (d-1) ~r_h & d=odd \\
                    \pi (d-2) ~r_h^2 & d = even \\
            \end{cases}
    \end{align}
\end{subequations}
Clearly they are all free of the Lovelock order and the spacetime dimension except for the proportionality constant. This shows that the thermodynamics is indeed universal. \\

The thermodynamics of spherically symmetric black hole solutions in general Lovelock gravity has been worked out in detail in \cite{cai-therm} and it provides general expressions for determining the temperature and entropy. Explicity these are given by a series in terms of powers of the horizon radius $r_h$ as:

\begin{subequations}
    \begin{align}
        T & = \frac{\sum_{i=0}^N (d-2i-1)\alpha_ir_h^{-2i+2}}{4\pi r_h \sum_{i=1}{N}i\alpha_i r_h^{-2i+2}}  \\
        S & = \frac{\Omega_d r_h^{d-2}}{4G} \sum_{i=1}^N\frac{i(d-2)}{d-2i} \alpha_ir_h^{-2i+2}
    \end{align}
\end{subequations}

For universality of thermodynamics we would now demand that the temperature and entropy  are always given in terms of the horizon radius as for the Einstein gravity in $3$ and $4$ dimensions. Their horizon radius dependence is entirely free of the spacetime dimension and the Lovelock order. This means for $\alpha_i \neq 0$ we must have
\be
    r_h^{d-2i} =    \begin{cases}
                r_h & d=odd \\
                r_h^2 & d=even
            \end{cases}
\ee
Thus $i = \lb\lfloor\frac{d-1}{2}\rb\rfloor = N$ and so the only terms that
contribute are $\alpha_0$ and $\alpha_N$; i.e. $\Lambda$ and the maximal order Lovelock. This is what characterizes the pure Lovelock black hole. This proves the sufficient condition that the universality uniquely singles out the pure Lovelock gravity. \\

We have thus proven that the necessary and sufficient condition for the thermodynamical universality of a static black hole is that it is the pure Lovelock black hole. The thermodynamical universality uniquely characterizes the pure Lovelock black hole. \\

Note that Eq (\ref{thepot}) will asymptotically go over to a
$d$-dimensional Einstein black hole in dS/AdS spacetime. This is
what should really happen because the high energy effects coming
from the higher order terms should die down asymptotically. This in
contrast to the dimensionally continued black hole that could never
go over asymptotically to the Einstein limit. At the high energy end
it however approaches the dimensionally continued black hole and
hence it has the desired behavior at that end too. Apart from the
universal thermodynamics, the pure Lovelock black hole has therefore
the expected asymptotic limits at both ultraviolet and infrared
ends.
\\.

It is remarkable that the thermodynamics as a function of the
horizon radius is thus completely insensitive and neutral to the
Lovelock order. This means the thermodynamics remains invariant
for the order of the Lovelock action. To understand why it happens, let us write
the potential in Eq.\eqref{thepot} as $V(r)=1-\Phi(r)$
\be
 \Phi = \lb(\alpha_0r^{2N} + \frac{\mu}{r^{d-2N-1}}\rb)^{1/N}
\ee For odd and even dimensions, $d-2N-1 = 0$ and $1$ respectively, which
makes the black hole potential the same as that for the Einstein
black hole ($N=1$) with a cosmological constant - in $3$ and $4$
dimensions respectively. For general $N$, the potential above is
essentially the $N$th root of it which gets squared out in the
definition of the horizon radius which is given by $\alpha_0 r^{2N}
+ \frac{\mu}{r^{d-2N-1}} =1$. That is the scaling of gravitational
potential with the Lovelock order is wonderfully compensated by the
scaling in the definition of the horizon
radius. \\

The important question is whether this universality is purely a
classical result or also true for quantum mechanical computations of
the entropy? Since we have computed entropy by simply integrating
the First Law of thermodynamics which is a general conservation law
and hence it should transcend to quantum computations as well. We
should therefore expect that the the universality should be true in
general, irrespective of classical or quantum computations. It would
indeed be very interesting to verify it by actually counting the
quantum mechanical degrees of freedom either following the lines of
the seminal work \cite{Strominger:1996sh} in the string theory
framework or using the loop quantum gravity approach as in
\cite{Asht}. We believe that universality would carry over to the
quantum calculations as well. If firmly established, this would
indeed be a very significant and important result for the black hole
thermodynamics as well as for the Lovelock gravity.
It is very significant that the thermodynamical universality characterizes the pure Lovelock black hole.
This would strongly suggest that the extension of the Einstein gravity for bringing in the high energy
effects has perhaps to follow the Lovelock way. \\

Finally probing of the universal features of gravity in higher
dimensions is a very pertinent and interesting question on its own
account. This enquiry began with the discovery that gravity inside a
uniform density sphere \cite{dmk} has the universal character in the
sense that it does not matter whether it is the Newton or Einstein
or in general the Lovelock gravity. In particular, it is always
described by the Schwarzschild interior solution irrespective of
whether it is the Einstein or Einstein-Lovelock theory in higher
dimensions. Such universal features may prove enlightening and
insightful in understanding the intricate working of gravity and its
higher dimensional dynamics. And the higher dimensions are tied to
the high energy effects of gravity \cite{LL}. For going beyond
Einstein we have to find a guiding principle like the Principle of
Equivalence which has also to have the universal character. We
believe that the present study has taken this effort to a very
significant step forward.

\section*{Acknowledgements}
JMP and KP thank IUCAA for warm hospitality for their separate
visits during which a significant part of this work was done.



\begin{thebibliography}{99}
\bibitem{lovelock} D. Lovelock, J. Math. Phys. {\bf 12}, 498 (1971); {\bf 13}, 874 (1972)
\bibitem{string} M. J. Duff, B. E. W. Nilsson and C. N. Pope, Phys. Lett. B {\bf 173}, 69 (1986); D. J. Gross and E. Witten, Nucl. Phys. {\bf B277}, 1 (1986); B. Zumino, Phys. Rep. {\bf 137}, 109 (1985); B. Zwiebach,Phys. Lett. {\bf 156B}, 315 (1985); D. Friedan, Phys. Rev. Lett. {\bf 45} (1980) 1057; I. Jack, D.Jones and N. Mohammedi, Nucl. Phys. {\bf B322}, 431 (1989); C. Callan, D. Friedan, E. Martinec and M. Perry, Nucl. Phys. {\bf B262}, 593 (1985).
\bibitem{dad} N. Dadhich, \emph{Universality, gravity, the enigmatic $\Lambda$ and beyond}, \arxiv{gr-qc/0405115}; \emph{Probing universality of gravity}, \arxiv{gr-qc/0407003}; \emph{On the Gauss Bonnet Gravity}, Mathematical Physics, (Proceedings of the 12th Regional Conference on Mathematical Physics), Eds., M.J. Islam, F. Hussain, A. Qadir, Riazuddin and Hamid Saleem, World Scientific, 331; (\arxiv{hep-th/0509126}); \emph{On Lovelock vacuum solution}, \arxiv{1006.0337}
\bibitem{Dadhich:2008df} N.~Dadhich,
  Pramana {\bf 74}, (2010) 875;
  (\arxiv{0802.3034}).
\bibitem{Dadhich:2010} N.~Dadhich and J.M. Pons,
  ``Consistent Levi Civita truncation uniquely characterizes the Lovelock Lagrangians.'' \arxiv{1012.1692}.
\bibitem{wheeler} J. T. Wheeler, Nucl. Phys. {\bf B268}, 737 (1986); {\bf B273}, 732 (1986).
\bibitem{whitt} B. Whitt, Phys. Rev. {\bf D38}, 3000 (1988).
\bibitem{dad-const} N. Dadhich, Pramana {\bf 77}, 1 (2011), (\arxiv{1006.1552}); \emph{On the measure of spacetime and gravity}, \arxiv{1105.3396}.
\bibitem{BTZ} M. Ba\~{n}ados, C. Teitelboim, and J. Zanelli, Phys. Rev. {\bf D49}, 975 (1994).
\bibitem{CTZ} J. Crisostomo, R. Troncoso, and J. Zanelli, Phys. Rev. {\bf D62}, 084013 (2000).
\bibitem{LL} N. Dadhich, Math. Today {\bf 26}, 37 (2011), (\arxiv{1006.0337}).
\bibitem{kpd} K. Prabhu, J. Pons and N. Dadhich, \emph{On some interesting properties of the Lovelock black holes}, under preparation
\bibitem{dmk} N. Dadhich, A. Molina and A. Khugaev, Phys. Rev {\bf D81}, 104026 (2010).
\bibitem{jac-my} T. Jacobson and R. C. Myers, Phys. Rev. Lett. {\bf 70}, 3684, (1993); (\arxiv{hep-th/9305016v1}).
\bibitem{cai-therm} R. G. Cai, Phys. Lett. {\bf B582}, 237, (2004);
(\arxiv{hep-th/0311240v2}).
\bibitem{Strominger:1996sh}
  A.~Strominger and C.~Vafa,
  Phys.\ Lett.\  B {\bf 379} (1996) 99
  \arxiv{hep-th/9601029}
\bibitem{Asht} A.Ashtekar, J.Baez, A.Corichi, K.Krasnov, Phys. Rev. Lett. {\bf 80}, 904,  (1998);
(\arxiv{gr-qc/9710007})
\end{thebibliography}
\end{document}